\documentclass[Physsubmission, Phys]{SciPost}

\hypersetup{
    colorlinks,
    linkcolor={red!50!black},
    citecolor={blue!50!black},
    urlcolor={blue!80!black}
}

\usepackage[bitstream-charter]{mathdesign}
\usepackage{bm}
\usepackage{subcaption}

\DeclareSymbolFont{usualmathcal}{OMS}{cmsy}{m}{n}
\DeclareSymbolFontAlphabet{\mathcal}{usualmathcal}

\begin{document}

\begin{center}{\Large \textbf{A new method to extract the valence transversity distributions \\
}}\end{center}

\begin{center}
Mauro Anselmino\textsuperscript{1},
Raj Kishore\textsuperscript{2$\star$} and
Asmita Mukherjee\textsuperscript{2}
\end{center}

\begin{center}
{\bf 1} Università degli Studi di Torino
\\
{\bf 2} Indian Institute of Technology Bombay
\\
* raj.theps@gmail.com
\end{center}



\definecolor{palegray}{gray}{0.95}
\begin{center}
\colorbox{palegray}{
  \begin{tabular}{rr}
  \begin{minipage}{0.1\textwidth}
    \includegraphics[width=22mm]{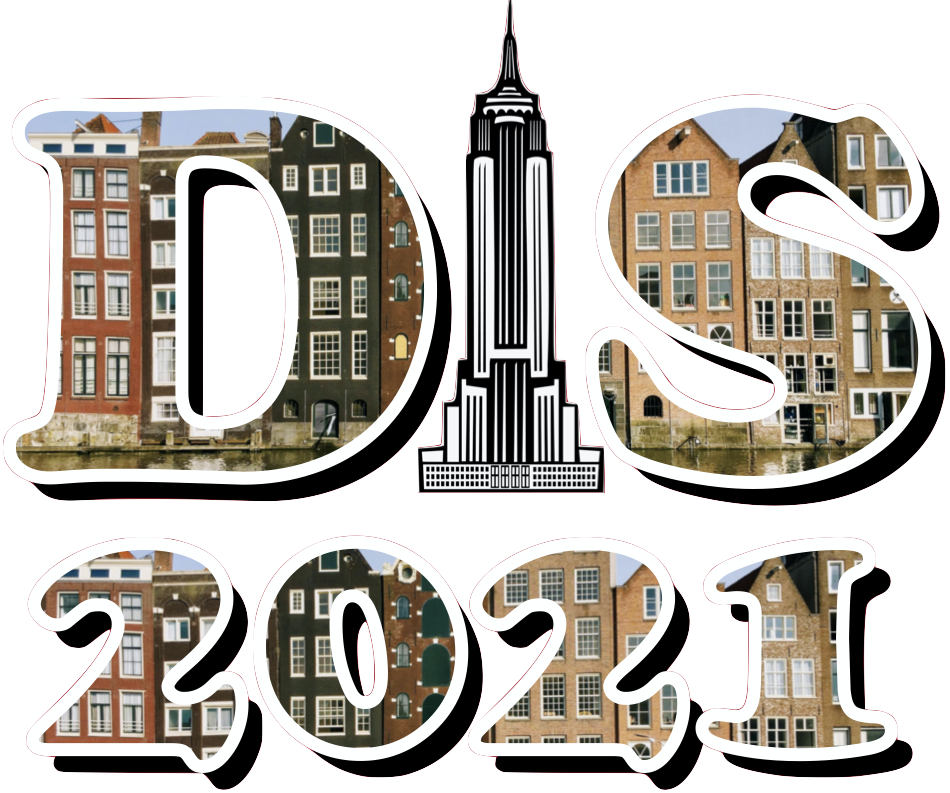}
  \end{minipage}
  &
  \begin{minipage}{0.75\textwidth}
    \begin{center}
    {\it Proceedings for the XXVIII International Workshop\\ on Deep-Inelastic Scattering and
Related Subjects,}\\
    {\it Stony Brook University, New York, USA, 12-16 April 2021} \\
    \doi{10.21468/SciPostPhysProc.?}\\
    \end{center}
  \end{minipage}
\end{tabular}
}
\end{center}

\section*{Abstract}
{\bf 
A new method is suggested for the extraction of $u$-quark and $d$-quark transversity distributions, using single spin asymmetry (SSA) data in semi-inclusive deep inelastic scattering (SIDIS) processes, where they couple to the Collins or the di-hadron fragmentation functions. We discuss a recent suggestion to extract the transversity distribution using the concept of difference asymmetries and their ratios, which avoids the requirement of Collins function. We suggest new measurements, involving ratios of polarized cross-sections, that would directly probe the ratio $h_1^{d_v}/h_1^{u_v}$. We also show some numerical estimates.
}


\section{Introduction and formalism}
\label{sec:intro}
In order to understand the quark structure of a polarized nucleon, the transversity distribution function, $h_{1}(x)$ or $\Delta_{T}q(x)$, contains fundamental and necessary information. $h_{1}(x)$, at the moment the least known, together with the unpolarized distribution functions $q(x)$ or $f_1(x)$ and the helicity distribution functions $\Delta q(x)$, would provide a full description of the polarized nucleon in the collinear ($\bm{k}_{\perp}$ integrated) configuration. So far, we don't have much information on $h_1(x)$ and the main reason is that due to its chiral-odd nature, it decouples and can not occur alone in DIS processes. The only way it can be accessed, in DIS processes, is by coupling this distribution with another chiral-odd function. Single spin asymmetry (SSA) in a SIDIS process, convolute $h_1(x)$ with, another chiral-odd function, called Collins fragmentation function \cite{Collins:1992kk}. In  Refs.~\cite{Anselmino:2007fs,Anselmino:2008jk,Anselmino:2013vqa,Kang:2015msa}, $h_1^u(x)$ and $h_1^d(x)$ have been extracted for the first time, using this process. Similar results were obtained by coupling the transversity distribution with a di-hadron fragmentation function~\cite{Bacchetta:2011ip,Bacchetta:2012ty,Radici:2015mwa}. 


We consider the SIDIS production of hadron $h$, $l+P^{\uparrow}\rightarrow l'+h+X$. We take virtual photon and proton center of mass frame, where $\gamma^*$ and proton are colliding along the $z$-axis with their respective momenta $\bm{q}$ and ${\bm{P}}$. The leptonic plane coincides with $xz$ plane. The initial quark momentum inside the parent target proton is $\bm{k}=x\bm{P}+\bm{k}_{\perp}$, where $\bm{k}_{\perp}$ is their intrinsic transverse momentum. The final hadron $h$ is produced from the fragmentation of the scattered quark. Its momentum $\bm{P}_h$, is given as $\bm{P}_h=z\bm{k'}+\bm{p}_{\perp}$, where $\bm{p}_{\perp}$ is the transverse momentum of the final observed hadron with respect to the direction of fragmenting quark (with momentum $\bm{k'}$). Following Refs.~\cite{Bacchetta:2006tn,Anselmino:2011ch}, in the case of a transversely polarized target colliding with an unpolarized lepton beam, the cross-section for the process is given as

\begin{eqnarray}
\frac{d\sigma^{\ell p(S_T)\to \ell^\prime h\,X}}
{dx_{B}\, dQ^2 \, dz_h \, d^2\bm{P}_{T} \, d \phi_S} &=&
\frac{2\alpha^2}{Q^4}\,\Bigl\{ \frac{1+(1-y)^2}{2}\,F_{UU} + \dots
\label{eq:dsig-sidis} \\
&+& \Bigl[ 
(1-y)\,\sin(\phi_h+\phi_S)\,F_{UT}^{\sin(\phi_h+\phi_S)} + \dots\,\Bigr]\Bigr\}\,,
\nonumber
\end{eqnarray}
where we have considered only terms which are relevant to the Collins asymmetry. The SIDIS variables $x_B$, $y$, $z_h$ and $Q$ have their usual meaning. $\phi_h$ and $\phi_S$ are the azimuthal angles, with respect to the leptonic plane, of the observed hadron and the spin polarization vector respectively. In Eq.~(\ref{eq:dsig-sidis}), $F_{UU}$ and $F_{UT}$ are the structure functions where the subscript $UU$ refers to the unpolarized case and $UT$ refers to an unpolarized beam with a transversely polarized target nucleon. 

The contribution to the single spin asymmetry, as defined in Eq.~(2) of Ref.~\cite{PhysRevD.102.096012}, from the Collins effect can be given as 

\begin{eqnarray}
A_{UT}^{\sin(\phi_h + \phi_S)} =
\frac{2(1-y)\,F_{UT}^{\sin(\phi_h+\phi_S)}}
{\bigl[ 1+(1-y)^2\bigr]\,F_{UU}} \equiv D_{NN} 
\frac{F_{UT}^{\sin(\phi_h+\phi_S)}}{F_{UU}}
\label{eq:AUT-coll}
\end{eqnarray}
where $D_{NN} = 2(1-y)/[1+(1-y)^2]$ is the quark depolarisation factor. Following Ref.~\cite{Anselmino:2011ch}, these structure functions can be expressed in terms of Gaussian parameterization for the TMDs in which $F_{UU}\propto f_{q/p}(x)\otimes D_{h/q}(z)$ and $F_{UT}^{\sin(\phi_h + \phi_S)}\propto h_1^q(x)\otimes\Delta^N D_{h/q^\uparrow}(z)$, where $f_{q/p}$ and $D_{h/q}$ represent collinear PDFs and FFs respectively, while $\Delta^N D_{h/q^\uparrow}(z)$ is the $z$-dependent part of the Collins fragmentation functions and $h_1^{q}(x)$ is the transversity distribution.
%
%
\section{A suggested measurement using difference asymmetries}

The authors of Ref.~\cite{Barone:2019yvn} have suggested a method, using the concept of difference asymmetry, $A_D$, for charged hadrons, to probe the transversity distributions. Following Ref.~\cite{Barone:2019yvn}, the polarized SIDIS cross-section given in Eq.~(\ref{eq:dsig-sidis}), can be written as
\begin{eqnarray}\label{csts}
\sigma_t^{\pm} = \sigma_{0,t}^{\pm} + \sin(\phi_h+\phi_S)\,
D_{NN} \, \sigma_{C,t}^{\pm} 
+ \dots 
\end{eqnarray} 
where the subscript $t$ indicates the type of target ($p$ for proton, $n$ for neutron and $d$ for deuteron) and the superscript $+$ or $-$ refers to observed $\pi^{+}$ and $\pi^{-}$. The expressions of $\sigma_0$ and $\sigma_C$ are obtained as  
\begin{eqnarray}
\sigma_0 = \frac{\alpha^2}{Q^4} \, \bigl[ 1+(1-y)^2\bigr]\,F_{UU}
\quad\quad\quad
\sigma_C = \frac{\alpha^2}{Q^4} \, \bigl[ 1+(1-y)^2\bigr] \,
F_{UT}^{\sin(\phi_h + \phi_S)}
\end{eqnarray}
A measure of the Collins asymmetry is taken as the ratio \cite{Barone:2019yvn}
\begin{eqnarray}\label{colasy}
A_C = \frac{\sigma_C}{\sigma_0} = \frac{F_{UT}^{\sin(\phi_h + \phi_S)}}{F_{UU}}
= \frac{1}{D_{NN}} \, A_{UT}^{\sin(\phi_h + \phi_S)}
\end{eqnarray}    
which differs by $1/D_{NN}$ factor, from the asymmetry, $A_{UT}^{\sin(\phi_h + \phi_S)}$, measured in the  experiments. 

The difference asymmetries for the charged hadron (pions) production is defined as \cite{Barone:2019yvn}
\begin{eqnarray}
A_{D,t} \equiv \dfrac{\sigma_{C,t}^+ - \sigma_{C,t}^-} 
{\sigma_{0,t}^+ + \sigma_{0,t}^-} = 
\dfrac{\sigma_{0,t}^+}{\sigma_{0,t}^+ + \sigma_{0,t}^-} \, A_{C,t}^+ - 
\dfrac{\sigma_{0,t}^-}{\sigma_{0,t}^+ + \sigma_{0,t}^-} \, A_{C,t}^- \>,
\label{adt}
\end{eqnarray}
where the 2nd equality can be used to extract $A_D$'s from the available data on the Collins
symmetry and the unpolarized cross sections.

In Ref.~\cite{Barone:2019yvn}, it is suggested to measure a ratio of difference asymmetries, which cancels out the dependence on the Collins function and remains with a ratio of transversity distributions along with PDFs and FFs;
\begin{eqnarray}
R_{D,d/p} \equiv \dfrac{A_{D,d}}{A_{D,p}} = 3 \left[ \dfrac
{(4f_1^{u} + 4f_1^{\bar u} + f_1^{d} + f_1^{\bar d})\,(D_{1,fav} + D_{1,dis})
	+ 2(f_1^{s} + f_1^{\bar s})\,D_{1,s}}
{5(f_1^{u} + f_1^{\bar u} + f_1^{d} + f_1^{\bar d})\,(D_{1,fav} + D_{1,dis})
	+ 4(f_1^{s} + f_1^{\bar s})\,D_{1,s}} \right]
\dfrac{h_1^{u_v} + h_1^{d_v}}{4h_1^{u_v} - h_1^{d_v}} \> \cdot \label{eq:ratio}
\end{eqnarray}
Assuming that one knows the unpolarized PDFs and FFs,  Eq.~(\ref{eq:ratio}) allows to direct access the ratio of valence $u$ and $d$ transversity distributions.  But, in this approach, the difference asymmetries are small quantities with large relative errors that could cause a huge uncertainty in their ratios, as shown also in Ref.~\cite{Barone:2019yvn} and~\cite{Compass:2020}. 

\section{A new measurement and numerical estimates}
\label{sec:another}
The new measurement~\cite{PhysRevD.102.096012} is motivated by the possibility to measure the $\sin(\phi_h+\phi_S)$ modulation of the SIDIS cross section (\ref{eq:dsig-sidis}), which relates to $\sigma_C$, Eq.(\ref{csts}), for different targets and observing both positively and negatively charged final pions. We can built a ratio of difference of polarized cross sections as 
\begin{eqnarray}\label{eq:newratio}
R_{C,d/p} \equiv \frac{\sigma_{C,d}^+ - \sigma_{C,d}^-}
{\sigma_{C,p}^+ - \sigma_{C,p}^-} = 3 \, \frac{h_1^{u_v} + h_1^{d_v}}{4h_1^{u_v} - h_1^{d_v}} 
= 3\, \frac{1 + \frac{h_1^{d_v}}{h_1^{u_v}}} {4 - \frac{h_1^{d_v}}{h_1^{u_v}}} 
\end{eqnarray}   
which has a simple partonic interpretation in terms of transversity distributions only. A similar expression can be obtained for the neutron target as well. Not only the Collins function but the PDFs and FFs also cancel out in ratio. Moreover, $R_{C,d/p}$ is not the ratio of two small quantities, unlike in the difference asymmetry approach. Hence, the ratio $R_{C,d/p}$ can have much smaller uncertainties.
%
%
\begin{figure}[]
	\begin{center}
		\begin{subfigure}{.48\textwidth}
			\includegraphics[width=7.8cm,height=3.9cm,clip]{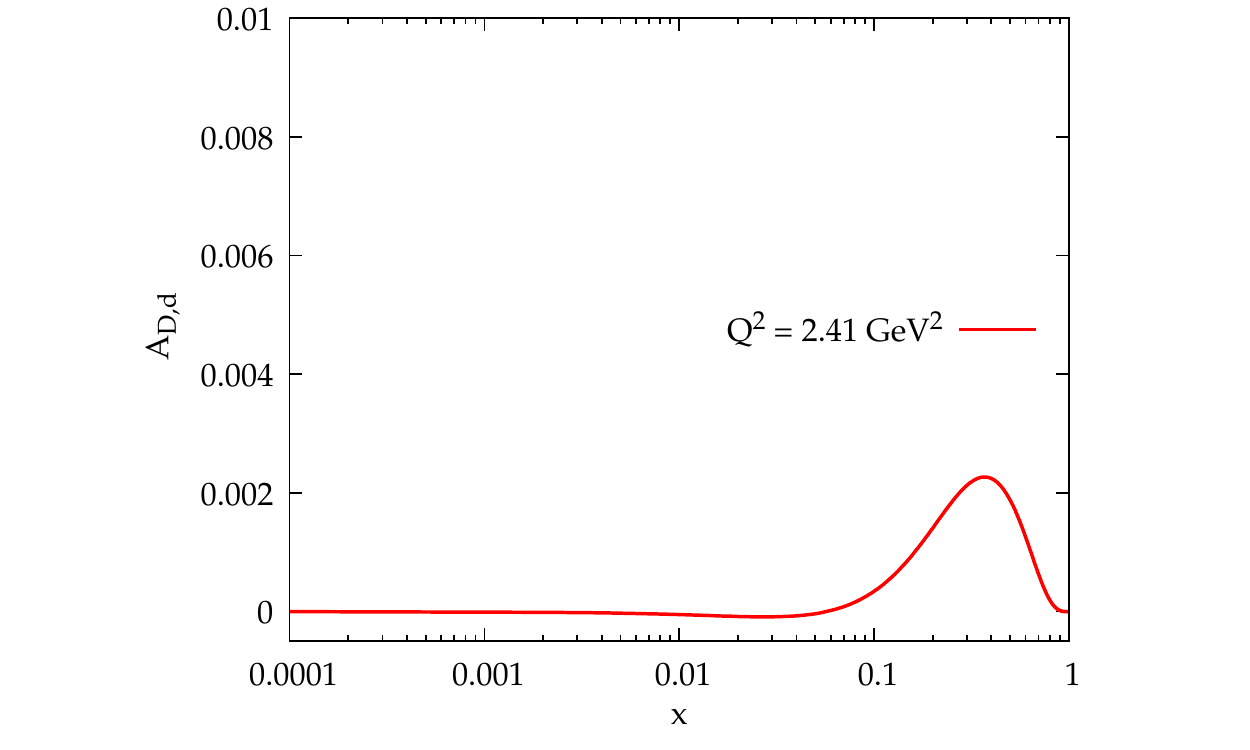}
		\end{subfigure}
		\begin{subfigure}{.48\textwidth}
			\includegraphics[width=7.8cm,height=3.9cm,clip]{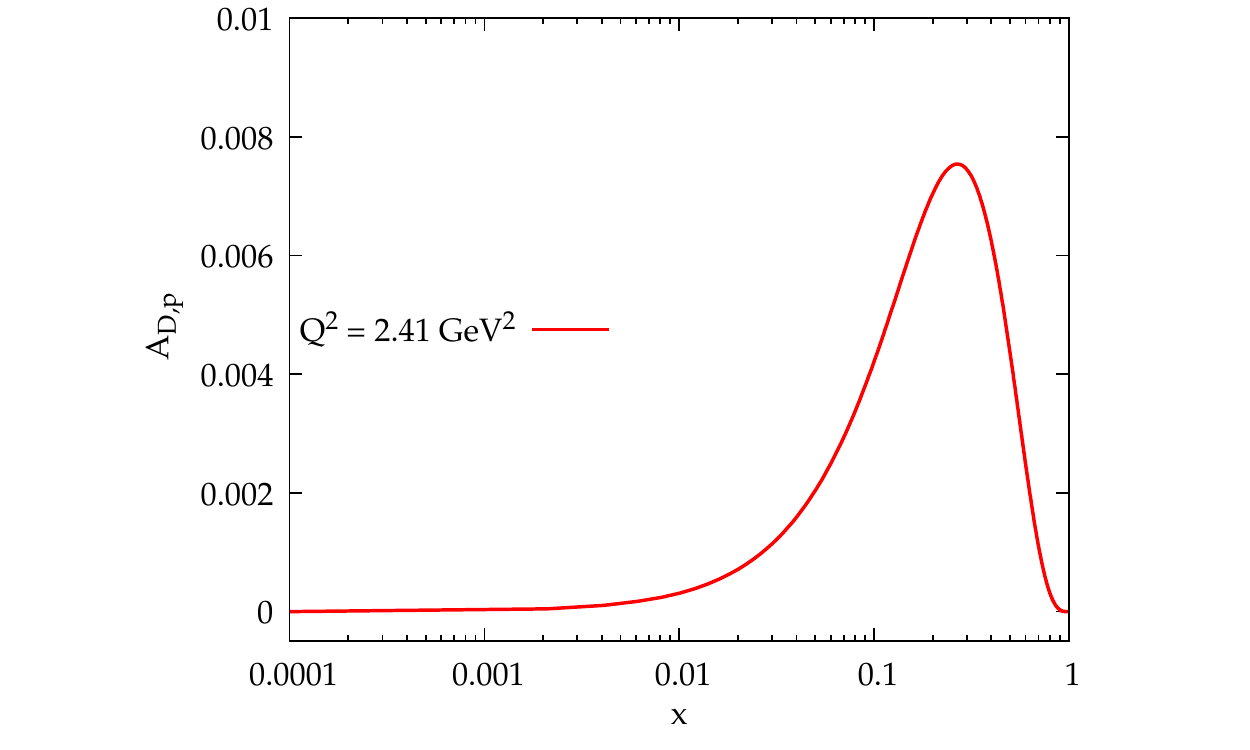}
		\end{subfigure}
	\end{center}
	\caption{\cite{PhysRevD.102.096012} \label{fig1} We plot the difference asymmetries (a) $A_{D,d}$ 
		(deutron target) and (b) $A_{D,p}$ (proton target) vs.~$x$ at $Q^2 = 2.41$ GeV$^2$. 
		We integrate the variables $z$ and $P_T$ over the ranges $0.1 < z < 1$ 
		and $0 < P_T < 5$ GeV.}
	\label{figure1}
\end{figure} 

For the numerical estimates, we use the parameterization and the best-fit parameters, for the transversity distributions and the Collins fragmentation functions, from Ref.~\cite{Anselmino:2013vqa}. MSTW2008 PDFs ~\cite{Martin:2009iq} are used and the unpolarised pion FFs are taken from Ref.~\cite{PhysRevD.91.014035}; we refer to Ref.~\cite{PhysRevD.63.094005} for the helicity distributions. For a description of the plots see the captions and Ref.~\cite{PhysRevD.102.096012}. 

\begin{figure}[]
	\begin{center}
		\begin{subfigure}{.48\textwidth} 
			\includegraphics[width=7.8cm,height=3.9cm,clip]{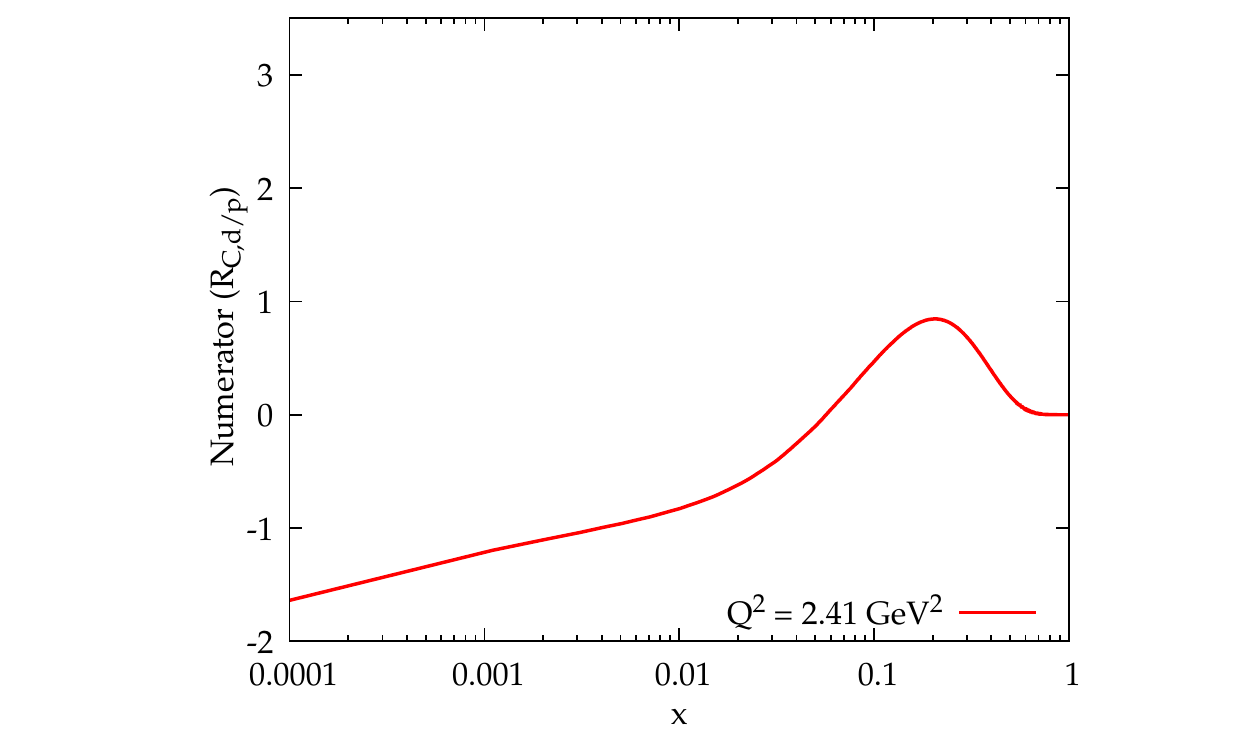}
		\end{subfigure}%
		\begin{subfigure}{.48\textwidth}
			\includegraphics[width=7.8cm,height=3.9cm,clip]{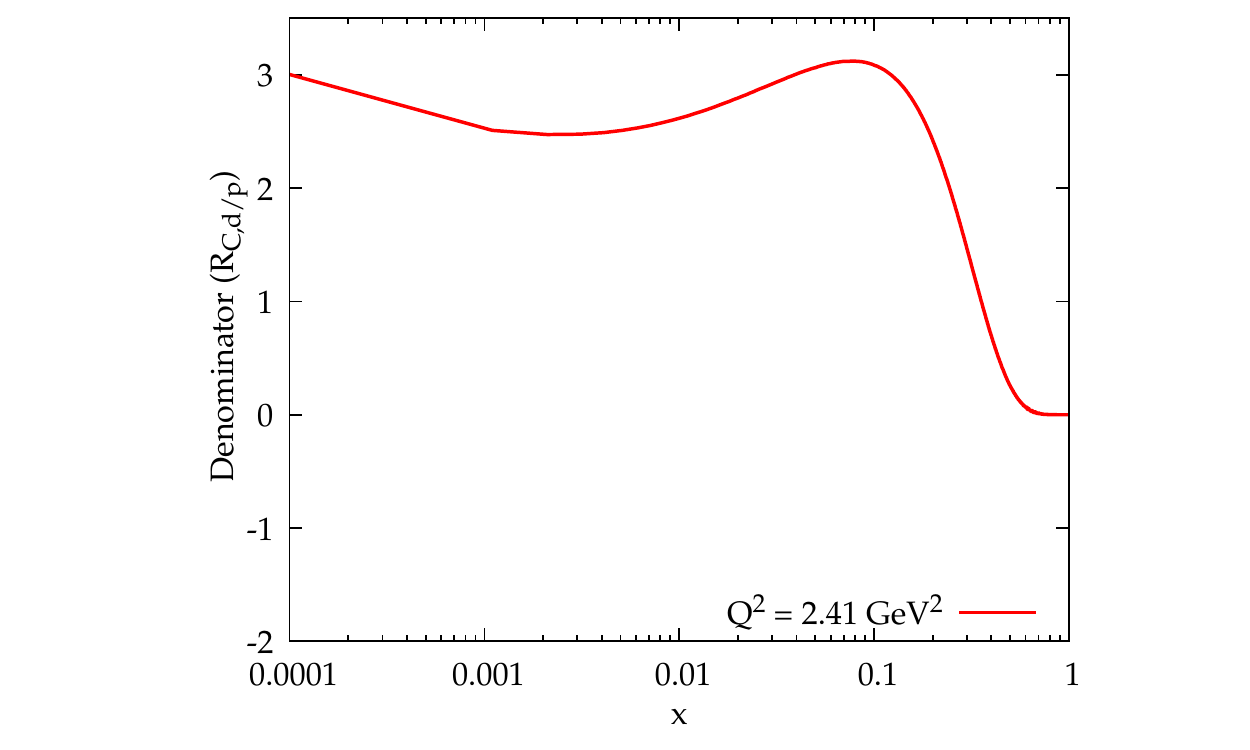}
		\end{subfigure}
	\end{center}
	\caption{\cite{PhysRevD.102.096012} \label{fig2} 
		We plot the numerator (a) and denominator (b) of 
		$R_{C,d/p}$  vs.~$x$ at $Q^2 = 2.41$. We integrate the $z$ variable over the range $0.1 < z < 1$.}
	\label{figure2}
\end{figure}

\section{Conclusion}
{\color{black} We see from the plots in Fig.~\ref{figure1}, that the difference asymmetries $A_D$'s, estimated from the existing knowledge of the Collins functions, the transversity distributions and the unpolarized PDFs and FFs are very small. Moreover, these estimates from SIDIS data have large errors and their ratio, although avoiding the knowledge of the Collins function, is bound to have huge uncertainties~\cite{Barone:2019yvn, Compass:2020}.}  

{\color{black} Our suggested asymmetries $R_{C,d/p}$ and $R_{C,n/p}$ are ratios of much larger quantities, which can be obtained from the $\sin(\phi_h+\phi_S)$ modulation of transversely polarized SIDIS cross sections. It might be a challenging measurement, but its TMD interpretation is much simpler, (Eq. \ref{eq:newratio}), and it directly probes the ratio  
$h_1^{d_v}/h_1^{u_v}$. Such a simplicity of $R_{C,d/p}$ should encourage its measurement, which could be done at COMPASS and JLab 12 experiments or at the future EIC facility.} 

\section*{Acknowledgements}
RK thanks the organisers of the workshop on Deep-Inelastic Scattering (DIS)-2021 at Stony Brook, New York, for the opportunity to present this work.



\bibliography{DIS_proceeding_bib.bib}

\nolinenumbers

\end{document}